# Measurement of surface concentration of fluorophores using fluorescence fluctuation spectroscopy

A. Delon[1], J. Derouard[1], G. Delapierre[2] and R. Jaffiol[1]

(1) Laboratoire de Spectrométrie Physique (UMR CNRS-UJF 5588), Université Grenoble I

BP 87 38402 Saint Martin d'Hères cedex France

(2) CEA/DRT/LETI/DTBS, 38054 Grenoble

Abstract: Fluorescence fluctuation spectroscopy is applied to study molecules, passing through a small observation volume, usually subjected to diffusive or convective motion in liquid phase. We suggest that such a technique could be used to measure the areal absolute concentration of fluorophores deposited on a substrate or imbedded in a thin film, with a resolution of a few μm. The principle is to translate the solid substrate in front of a confocal fluorescence microscope objective and to record the subsequent fluctuations of the fluorescence intensity. The validity of this concept is investigated on model substrates (fluorescent microspheres) and DNA biochips.

Fluorescence correlation spectroscopy (FCS) consists in calculating the autocorrelation of the fluorescence intensity time trace emanating from the small confocal volume illuminated by a focused laser beam.[1] From the analysis of this correlation function one is able to get information about the kinetics of the fluorescent molecules (photochemical processes, transport, interactions) and about the average number of fluorescent molecules in the observed volume, that is about their absolute concentration. This information is in many cases of great interest and often very difficult to get from other methods.

A complementary method to FCS is based on the analysis of the moments of the statistical distribution of photon counts or of the photon count histogram.[2-4] It has been shown that, in some conditions it is possible using this method to identify the presence of two species which differ by their brightness and to measure their absolute concentrations. The corresponding experiments have been carried out in solution, liquid membranes, or in biological cells, where the molecules are spontaneously moving due to diffusion. In the present work we consider the case of fluorophores anchored on a solid substrate, translated with respect to the laser beam. Like in the case of solutions, the concentration of fluorophores can, in principle, be extracted from these fluctuations. The aim of this work is to investigate experimentally the validity of this concept. Potential applications include the characterization of fluorescent biochips or the measurement of fluorescent defects in optical thin films.

An alternative method is image correlation spectroscopy,[5] which consists in calculating the autocorrelation function of the confocal image. An analogous approach, recently applied to solid substrates, consists in averaging the autocorrelation functions corresponding to the successive lines of the confocal scan,[6] reviving the original scanning-FCS (S-FCS) technique initially developed by Petersen in the 1980s.[7] Other recent versions of S-FCS, consisting in moving the laser beam in respect to the sample, are proposed to access both spatial and temporal information of biological media.[8,9].

Two kinds of samples, deposited on a glass coverslip, have been studied: 20 nm fluorescent nanospheres (FluoSpheres, Invitrogen) and Rh6G labeled DNA (OliGold, Eurogentec,). The nanospheres samples were prepared by evaporating a drop of a diluted suspension of nanospheres in water (at pH 10, to prevent their aggregation). For the DNA samples, spots of single strand DNA molecules, 21 bases long, were deposited from a highly diluted solution. A small liquid chamber was placed onto it and the DNA were hybridized with the complementary strand labeled by Rh6G. Our experimental set up consists in a home made confocal microscope built from an inverted microscope. The 488nm radiation from an air cooled Ar+ laser was directed to the sample through a water immersion objective (UPLAPO 60× NA = 1.2, Olympus). The excitation power at the sample was about 1µW (nanospheres) or 100µW (Rh6G labeled DNA chips). The emitted fluorescence was detected with an avalanche photodiode (PerkinElmer). A home-made data acquisition system recorded the delay time between consecutive photons while moving the sample with a piezoelectric device (Piezosystem Jena). The 2D observation volume, depends on the size and shape of the focused laser beam onto the sample and on the geometry of the collection of the fluorescence. It can be approximated by a gaussian function, $W(r) = \exp(-2r^2/w_0^2)$, where the value of $w_0$ is extremely sensitive to the position of the observed sample surface with respect to the focal plane of the objective.

Fig. 1(a) shows the photon count rate, $I_{fluo}$, recorded when a fluorescent nanosphere sample is translated at constant velocity ($v$=10µm/s). The surface concentration of the nanospheres is low enough so that single fluorophores can be detected as isolated peaks. Their width corresponds to $w_0 \cong 0.25 \mu m$. Several facts can account for the dispersion of the peaks height: dispersion of the brightness of the nanospheres, formation of aggregates, and position of each nanosphere with respect to the center of the focal spot.

Fig. 1(b) shows the corresponding autocorrelation functions for different translation velocities $v$. The amplitude of the autocorrelation function may vary from one scan to another due to slight uncontrolled displacements of the sample holder along the optical axis. Therefore, for an easier comparison of their shapes, the amplitudes of the curves in Fig. 1(b) have been renormalized to their mean value. These autocorrelation functions show a plateau whose duration, $\tau_c$, is the time it takes for the focal spot to move on the sample by about its radius. Actually, as expected, the autocorrelation functions can be fairly well fitted by a gaussian function: [10]

$$G(\tau) = 1 + [G(0) - 1] \times \exp[-(\tau/\tau_c)^2] \quad (1)$$

The product $v\tau_c$ is in the range 0.3 - 0.4 μm, close to the value $w_0$ estimated above from the width of the fluorescence peaks. The value at the plateau, $G(0)$, gives the mean effective number $<N>$ of fluorescent particles inside the observation volume, assumed to be a 2D gaussian:

$$<N> = \frac{\gamma_2}{G(0) - 1} \quad (2)$$

where the shape factor $\gamma_2 = \frac{1}{2}$ and from which we determine the surface concentration of the fluorophores $C = 2<N>/\pi w_0^2$.[4]

With $G(0) = 5.9$ and $w_0 = 0.35\,\mu m$, we obtain $C = 0.53$ particles per $\mu m^2$, which is the order of magnitude found by directly counting the particles on an wide field image. Finally we note that the experimental curves differ from the theoretical ones by some oscillatory behavior which is fairly reproducible. These oscillation are due to mechanical vibrations of our experimental set up.

However, $<N>$, and thus $C$, can also be obtained from the two first moments of the statistical distribution of photon counts.[3] Let $k$ be the number of photons counted during each bin time $\delta t$. From the mean, $<k>$ and the variance, $<\Delta k^2>$, of $k$ estimated during time intervals of duration $\Delta t \gg \delta t$, one can derive the relation:[4]

$$<N> = \gamma_2 \frac{<k>^2}{<\Delta k^2> - <k>} \qquad (3)$$

We have checked numerically on our data, and it can be shown theoretically that this value of $<N>$ is identical to that of Eq. (2) where $G(0)$ is calculated over the same time interval $\Delta t$ ($\Delta t$ typically ranges from 0.1 to 0.5 s) provided that $\delta t \ll \tau_c$.[4] Finally, we stress the fact that calculating $<N>$ with the photon statistics avoids calculating the autocorrelation function and fitting it to determine $G(0)$.

In the situation corresponding to Fig. 1 the fluorophores are so diluted that they can be counted one by one. A more interesting situation occurs when the concentration is such that the individual particles cannot be resolved anymore. This is typically the case for fluorescent DNA biochips, as shown in Fig. 2. We see that there is an obvious and strong correlation between the photon count rate, $I_{fluo}$, and the effective, locally averaged, number of fluorescent molecules, $<N>$, deduced from the short time scale fluctuations of $I_{fluo}$. The statistical uncertainty on $<N>$ and $I_{fluo}$ can be estimated using the variances of the first two factorial cumulants, $\kappa_{[1]}$ and $\kappa_{[2]}$.[3] The statistical uncertainty on $I_{fluo}$ is negligible in our case.

Fig. 2 shows that the relationship between $I_{fluo}$ and $<N>$ is neither linear (the variations of $I_{fluo}$ are not proportional to those of $<N>$) nor single valued (two regions of the substrate having the same fluorescence intensity may have different $<N>$). Several facts may explain this discrepancy: i. When there are molecules of different brightness, Eq. (3) gives an effective number, $<N_{eff}>$, which can be associated to an effective brightness, $\varepsilon_{eff} = I_{fluo}/<N_{eff}>$, while the actual mean number of molecules $<N>$ is related to the mean of the brightness distribution $<\varepsilon> = I_{fluo}/<N>$. Actually a number of facts can affect the fluorescence quantum yield which may change from one molecule to the other and result in a distribution of brightness.[11] Following Petersen, $\varepsilon_{eff}$ can be expressed as a function of the mean, $<\varepsilon>$ and the variance, $<\Delta\varepsilon^2>$, of the brightness distribution:[7]

$$\varepsilon_{eff} = <\varepsilon>(1+ <\Delta\varepsilon^2>/<\varepsilon>^2) \qquad (4)$$

As a result, when $<\Delta\varepsilon^2>^{1/2}$ is much smaller than $<\varepsilon>$, Eq.(4) shows immediately that $\varepsilon_{eff} = <\varepsilon>$ and thus $<N_{eff}> = <N>$. However, when $<\Delta\varepsilon^2>^{1/2} > <\varepsilon>$ then $\varepsilon_{eff}$ is larger than $<\varepsilon>$ and $<N_{eff}>$ is smaller than the actual number of molecules $<N>$ by a factor which may depend on the position if the statistical distribution of $\varepsilon$ is position-dependent. ii. If the spatial distribution of the fluorophores over the distance $\Delta x = v \Delta t$ is not locally poissonian then Eq.(4) underestimates the actual value of $<N>$ due to the enhanced fluctuations of the fluorescence signal which result. iii. Other non-poissonian causes of fluctuations of the fluorescence signal, like the mechanical vibrations seen on Fig.1 have the same effects.

In conclusion, we have reported a novel method, based on fluctuation analysis, that maps the absolute concentration of fluorescent particles, anchored on a substrate, with a typical resolution of a few µm, and estimates the corresponding statistical uncertainties (performing FCS at the same resolution would provide a noisy and useless temporal information). The measurement may be biased due to several effects. If the S/N ratio is high enough the analysis of the higher moments of the fluorescence signal should in principle point out these effects. Finally, it must be emphasized that the method is neither sensitive to the intensity of the laser nor to the sensitivity of the photon detector and that it does not need a calibration, except the width of the "molecular detection efficiency function", $w_0$, which can be calculated from the width, $\tau_c$, of the autocorrelation function. The statistical analysis of photon counts can provide important information when studying fluorophores deposited on a surface, as encountered in biochips.

We gratefully acknowledge P. Barritault and S. Getin (CEA-LETI) for helpful discussions and encouragements.

**Fig. 1**

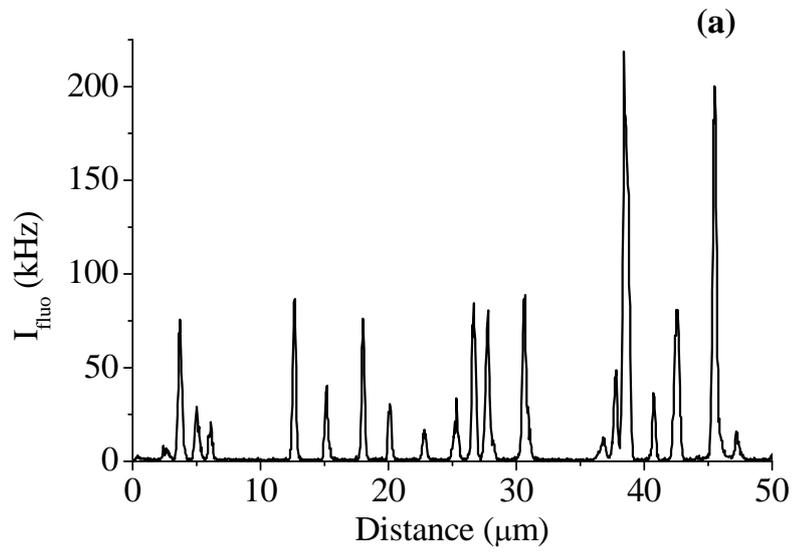

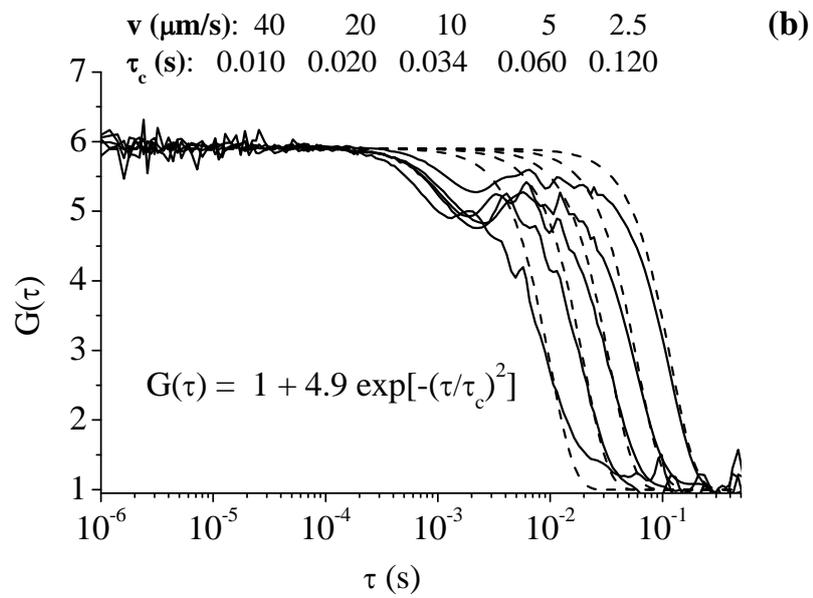

**Fig. 2**

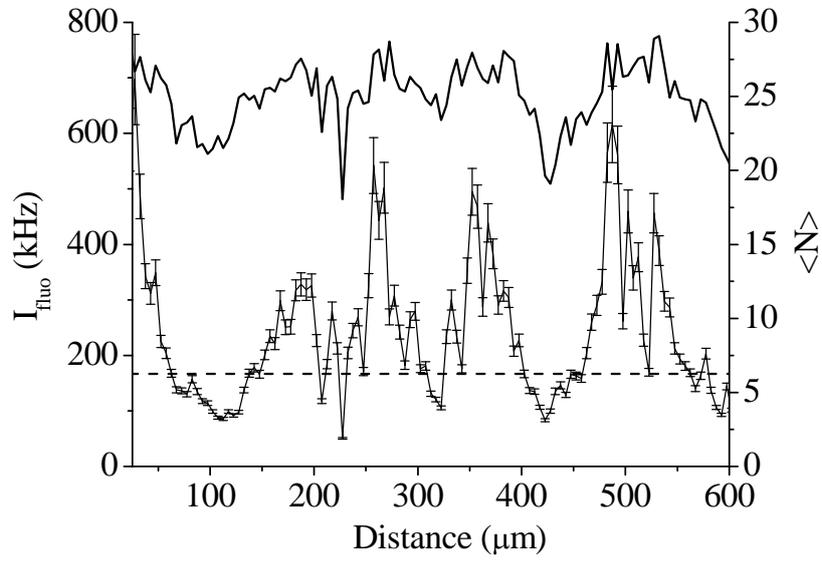

**Figure captions**

Fig. 1. Fluorescence data obtained by translating a sample containing highly diluted nanospheres; (a) fluorescence count rate *versus* the position; (b) autocorrelation functions obtained by translating the sample at various speeds, the solid lines are the experimental data (speeds decrease from left to right) and the dashed lines are the theoretical curves (Eq. (1)) with various characteristic times, $\tau_c$, given in the inset.

Fig. 2: Scan of fluorescent DNA molecules deposited on a glass coverslip. The solid line is the local count rate, averaged over distances of 5µm ($\Delta t$=0.1 s). The solid line with error bars is the local number of fluorescent molecules at the focal spot estimated from the fluctuations of the fluorescence using Eq. (4). Error bars are estimated according to Müller.[3] The horizontal dashed line is the number of molecules given by Eq. (2) with the autocorrelation function calculated over the whole scan.